\documentstyle[editedvolume]{crckapb}

\begin{opening}
\title{The Mildly Metal-Poor Globular Cluster M4:\protect\\
       Star-to-Star Abundance Variations in 20 Giants}

\author{I. I. Ivans}
\author{C. Sneden}
\institute{McDonald Observatory, University of Texas, Austin, USA}
\author{R.P. Kraft}
\institute{Lick Observatory, University of California, Santa Cruz, USA}

\end{opening}

\runningtitle{Mildy Metal-Poor M4 Star-to-Star Abundance Variations}

\begin{document}

Many low-metallicity globular clusters exhibit large star-to-star variations of C, N, O, Na, Mg, \& Al abundances.  In higher-metallicity clusters, the abundance swings are muted.  Evidence of anticorrelations of O versus Na and Mg versus Al exists in classical northern-hemisphere clusters spanning a range of metallicities, --0.8 $\le$ [Fe/H] $\le$ --2.2.  These and other anticorrelations, observed to be a function of giant branch position, are consistent with material having undergone proton-capture nucleosynthesis, presumably taking place in regions where the CNO-cycle is in operation, and brought to the surface by a deep-mixing mechanism.  Distinctly bimodal distributions of cyanogen strengths at nearly all giant branch positions are observed in some clusters, including  M4 (Norris 1981).  The metallicity of M4 ([Fe/H] $\sim$ --1) places it among clusters in which the O versus Na and Mg versus Al anticorrelations are expected to be suppressed. Indeed, not much abundance variation has been observed in small samples of M4 giant stars (see e.g. Brown \& Wallerstein 1992).  This puzzle led us to consider an abundance study of a large sample of bright giants in the mildly metal-poor globular cluster M4, the nearest, brightest, and one of the most accessible targets in which to study the CN-bimodal phenomenon. 

We observed 20 M4 giant stars in 1997 using high resolution \'echelle spectrographs covering the effective wavelength range: $\lambda\lambda$ 5200 to 8800${\rm \AA}$ on the 2.7-m at McDonald Observatory (R $\approx 60,000$) and the 3.0-m at Lick Observatory (R $\approx 50,000$).

Large and differential interstellar reddening due to the dark nebulosity in Scorpio-Ophiuchus does not permit a careful estimate of reddenings for individual stars; one cannot assume a strict $(B-V)$ versus Teff relationship.  Instead, we obtained precise Teffs by employing two different spectroscopic techniques: (1) ratios of central depths of temperature sensitive absorption features (Gray 1994) to initially estimate relative stellar Teffs and (2) as in previous Lick-Texas work, final Teffs were determined in conjunction with surface gravity and microturbulence constraints.

\begin{figure}[h]
\vspace{4.3cm}
\includegraphics{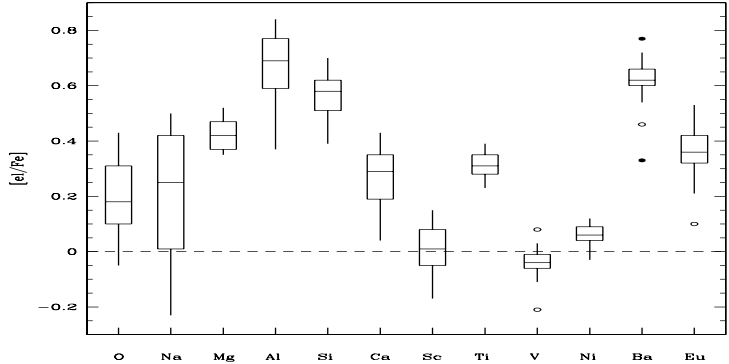}
\caption{Boxplot of M4 Giant Star Abundances - A boxed horizontal line indicates the interquartile range [IQ] \& median found for a particular element.  Vertical tails extending from the boxes indicate the total range of abundances determined for each element, excluding outliers.  Mild outliers (within 1.5$\times$IQ) are denoted by hollow circles (o) \& severe outliers (from 1.5 to 3$\times$IQ) by filled circles ($\bullet$).  The dashed line represents the solar value for a particular elemental abundance.}
\end{figure}

The scatter about the mean (of elements not expected to be sensitive to proton-capture nucleosynthesis) compares well to those obtained in other high resolution cluster work and, as expected, no trend with Teff is observed for these non-volatiles.  In individual M4 stars, anticorrelated abundances of N/Na/Al with C/O provide strong evidence for high-temperature proton-capture nucleosynthesis.  We also observe increasing O-depletion/N-enhancement as stars ascend the RGB but combined log$_{\epsilon}$(C+N+O) remains $\sim$ constant.  Extremely low carbon isotope ratios ($^{12}$C/$^{13}$C = 4.6 $\pm$ 0.8) indicate that the onset of mixing of the CNO-cycled material occurs at a luminosity lower than that of our observed stars.  Elements Na, Mg, Si, \& Ca are all enhanced with respect to the iron abundance, in approximate agreement with previous high resolution studies of M4 (Brown \& Wallerstein, 1992).  Al and Ba show uniform enhancements indicative of primordial enrichment.  The analysis also confirms anomalous absorption properties of dust along line of sight to M4 ($E(V-K)/E(B-V) = R = 3.8 \pm 0.4$) that deviate from the normal law of interstellar extinction.  


\end{document}